\documentclass[12pt,english,aps,manuscript]{article}
\usepackage[T1]{fontenc}
\usepackage[latin1]{inputenc}
\usepackage{geometry}
\usepackage{amsmath}
\usepackage{amssymb}
\usepackage[numbers]{natbib}

\makeatletter

\usepackage{geometry}

\makeatletter

\usepackage{color}

\makeatletter
\newcommand{\bee}{\begin{equation}}
\newcommand{\ee}{\end{equation}}
\newcommand{\beea}{\begin{eqnarray}}
\newcommand{\eea}{\end{eqnarray}}

\makeatother

\makeatother

\makeatother

\usepackage{babel}
\begin{document}
\begin{center}
\textbf{\Large AMSB and the Logic of Spontaneous SUSY Breaking }
\par\end{center}{\Large \par}

\begin{center}
\vspace{0.3cm}
 
\par\end{center}

\begin{center}
{\large S. P. de Alwis$^{\dagger}$ } 
\par\end{center}

\begin{center}
Physics Department, University of Colorado, \\
 Boulder, CO 80309 USA 
\par\end{center}

\begin{center}
\vspace{0.3cm}
 
\par\end{center}

\begin{center}
\textbf{Abstract} 
\par\end{center}

A cardinal principle of any theory of spontaneous (${\cal N}=1$)
supersymmetry breaking, is that the order parameter is a linear combination
of the F and D terms. Also as long as the supersymmetry of the action
is preserved at the quantum level, this principle should be valid
after the appropriate corrections are incorporated into the Kähler
potential, the superpotential, and the gauge coupling functions that
define the theory. The claim that in AMSB there is an extra term for
the gaugino mass that is proportional to the gravitino mass, is then
equivalent to the statement that quantum effects cause an explicit
breaking of (local) SUSY. Expanding on previous work, we argue that
this arises from confusing the scalar compensator with the density
compensator. We also comment on various recent papers on AMSB. 

\begin{center}
\vspace{0.3cm}
 
\par\end{center}

\vfill{}

$^{\dagger}$dealwiss@colorado.edu

\eject

\section{Introduction}

The order parameter of spontaneous supersymmetry breaking is a linear
combination of the F-terms of physical chiral superfields and the
D-terms associated with physical gauge pre-potentials. This is the
case both in global and local supersymmetry. In particular as is clear
from the component expansion of the general SUGRA Lagrangian given
for instance in \citep{Wess:1992cp}, the gaugino mass term can only
be generated if a physical chiral scalar field (or fields) acquire
a non-zero vacuum value. 

Now one may ask whether this general formula, wherein the entire (two
derivative) supersymmetric action is expressed in terms of three functions,
the real analytic Kähler potential $K$ the analytic superpotential
$W$ and the analytic gauge coupling function $f$, is valid at the
quantum level. The general expectation is that this is indeed the
case. In other words the quantum corrections are not supposed to change
this general structure. The only result of incorporating quantum effects
is to change the expressions for $K$ both at perturbative and non-perturbative
levels, and that for $W$ purely at the non-perturbative level, and
that for $f$ non-perturbatively as well as perturbatively to one
loop. This is essentially the content of the non-renormalization theorems.
Formulae which are completely consistent with this logic were presented,
for both the soft scalar mass terms, $A$ and $B$ terms, as well
as the gaugino mass, by Kaplunovsky and Louis (KL) in \citep{Kaplunovsky:1993rd,Kaplunovsky:1994fg}
several years before the first AMSB paper appeared.

In AMSB on the other hand a supersymmetry breaking term is present
in the expression for the gaugino mass that does not originate from
the F-term (or the D-term) of a physical field. Instead it is supposed
to come from the F-term of a conformal compensator field \citep{Randall:1998uk}\citep{Giudice:1998xp}.
After these papers there have been several \citep{Bagger:1999rd}\citep{Gripaios:2008rg}\citep{Conlon:2010qy}\citep{D'Eramo:2012qd},
that have attempted to provide justification for this. Regardless
of the details of these arguments none of these authors have made
any attempt to explain how their results can be reconciled with the
generally accepted logic of supersymmetry breaking and the earlier
result of KL.

The present paper is based on the following logic:
\begin{itemize}
\item Standard arguments in SUGRA lead to the conclusion that the gaugino
is a sum of terms each of which is proportional to the F-term of a
physical chiral scalar field. In particular if no F-term of such a
chiral scalar is non-zero, then the gaugino mass vanishes.
\item The above statement should be valid in the quantum theory as well
since the structure of the action (in terms of $K,W,f$), should remain
unchanged. All that can change is the expressions for these functions
in terms of the fields of the theory. 
\item Any argument that claims to find an extra source of (local) supersymmetry
breaking must therefore necessarily be claiming that the previous
statement is not correct. In other words these arguments must imply
that quantum effects create an explicit breaking of supersymmetry
- either in the Wilsonian effective action or the 1PI action or both.
\item The supergravity formalism does not allow for additional physical
effects coming from the conformal compensator. The physics of the
theory should not depend on the conformal compensator. In particular
one should be able to derive any physical effect from the superspace
supergravity formulation given in \citep{Wess:1992cp}.
\end{itemize}
For the sake of clarity let us expand upon these points. Let begin
with the last one. The starting point then is the superspace action
(\citep{Wess:1992cp} eqn (25.1)).

\begin{eqnarray}
S & = & -3\int d^{8}z{\cal E}(-\frac{\bar{\nabla}^{2}}{4}+2R)\exp[-\frac{1}{3}K(\Phi,\bar{\Phi};Q,\bar{Q}e^{2V})]+\nonumber \\
 &  & \left(\int d^{8}z{\cal E}[W(\Phi,Q)+\frac{1}{4}f(\Phi){\cal W}^{a}{\cal W}^{a}]+h.c.\right).\label{eq:actionC=00003D1}
\end{eqnarray}
This action is not however in the Einstein frame. In order to get
to this frame one must perform a (super) conformal transformation/field
redefinitions which in component form are given by eqns 21.15 to eqns
21.20 in \citep{Wess:1992cp} and are equivalently given by the superspace
transformations \eqref{eq:conf1}-\eqref{eq:conf4} (see below - next
section) with the transformation superfield $\tau$ fixed by 
\begin{equation}
2\tau+2\bar{\tau}=-\frac{\hat{K}(\Phi,\bar{\Phi})}{3}|_{H},\label{eq:weylgaugefix-1}
\end{equation}
as observed in KL. 

If these transformations had been non-anomalous (i.e. if the path
integral measure had been invariant) then the resulting action (in
component form after eliminating auxiliary fields) is given by eqn
(G.2) of \citep{Wess:1992cp} . However the transformations are anomalous
since they also involve transformations on chiral fermions. By general
arguments in the theory of chiral anomalies the only effect of these
(at two derivative level in the action) is to change the gauge coupling
function $f$ (superfield) by a term proportional to $\tau$. This
will then lead in particular to corrections to both the gauge coupling
and the gaugino mass. In the absence of an anomaly the latter would
be given by $-\frac{g^{2}}{4}F^{i}\partial_{i}f.$ In the presence
of the anomaly there would be an additional term proportional to $-\frac{g^{2}}{4}F^{\tau}\partial_{\tau}\tau$
where from the definition/fixing of the transformation superfield
\eqref{eq:weylgaugefix-1} $F^{\tau}=\frac{1}{6}F^{i}\partial_{i}\hat{K}$.
In particular this means that if the F-terms of the physical fields
$\Phi^{i}$ are zero in some vacuum (in particular supersymmetry is
unbroken) then the gaugino mass is zero. Any additional terms in the
formula for the gaugino mass can only come from an explicit breaking
of local supersymmetry.

In the next section we will outline the essential difference between
KL and the AMSB papers. Then we will show that the reason that the
AMSB literature appears to find an extra term is due to the misidentification
of a scalar compensator (which is what is correctly used in KL) with
a density compensator%
\footnote{See chapter 5 of \citep{Gates:1983nr} where the scalar compensator
is referred to as the tensor compensator - since in the non-minimal
case it is a non-zero rank tensor.%
}. The latter in fact cancels the Weyl anomaly at the expense of introducing
a diffeomorphism anomaly. Next we review the arguments given in various
papers to justify AMSB. 

Much of the discussion in this paper was essentially given four years
ago in \citep{deAlwis:2008aq}. However it was rather terse and in
any case has been widely ignored in the phenomenological literature.
Here we give an extended version of this discussion, in addition to
some new material.

\section{The Essence of the Difference}

The expression for the gauge coupling function has three anomalous
quantum components in addition to the classical contribution. The
first is the true Weyl anomaly which comes from transforming the supermetric
from the standard off-shell supergravity superspace supermetric to
the Einstein-Kähler frame metric. The second comes from transforming
the matter (e.g. MSSM) metric to a canonical metric - the so-called
Konishi anomaly. The third comes from transforming from supergravity
normalization of the gauge kinetic term to canonical normalization.
There is no discrepancy in the last two terms, between the generalized
AMSB formula as presented for instance in \citep{Bagger:1999rd} (which
we will call the BMP formula) and the original KL formula, and indeed
the corresponding contribution to the gaugino mass vanish when the
F-terms of physical fields are set to zero. The difference lies in
the Weyl anomaly term.

As mentioned in KL there are two ways of looking at the Weyl anomaly.
One way is to regard this as the effect of a field redefinition transformation
which is completely analogous to the one which results in the Konishi
or gauge kinetic term redefinition. In this case the corresponding
anomaly term (coming from the fact that the measure in the functional
integral is not invariant under the field redefinition) is \citep{Kaplunovsky:1994fg}
(see also our eqn.\eqref{eq:ftransform}),
\begin{equation}
f(\Phi)\rightarrow f(\Phi)+\frac{3c}{4\pi^{2}}\tau.\label{eq:ftransWeyl}
\end{equation}
Here $\tau$ is the chiral superfield transformation parameter and
$\Phi$ are the `moduli' fields which in principal may develop vacuum
expectation values as well as F-terms. The anomaly coefficient $c_{a}$
is given by (note that our definitions are the negatives of those
in KL)
\begin{equation}
c_{a}=T(G_{a})-\sum_{r}T_{a}(r)\label{eq:ca}
\end{equation}
 and $T(G_{a}),T_{a}(r)$ are the trace of a squared generator in
the adjoint and the matter representation $r$ of the gauge group
$G_{a}$. For future use we also give here the 1-loop $\beta$-function
coefficient
\begin{equation}
b_{a}=3T(G_{a})-\sum_{r}T_{a}(r).\label{eq:beta}
\end{equation}
In order to get to the Einstein-Kähler gauge $\tau$ has to be fixed
by the relation \eqref{eq:weylgaugefix-1}.

The instruction on the RHS requires one to take the harmonic part
of the real superfield $K$ and is essentially a Wess-Zumino gauge
fixing. It should be emphasized that the above relation is simply
the superspace form of the set of field redefinitions that are done
(see for example \citep{Wess:1992cp}) to get to the Einstein-Kähler
frame. Here $\hat{K}$ is the Kähler potential of the `moduli' fields
see eqn.\eqref{eq:Kexpn}. Taking the F-term of this gives the correction
to the classical gaugino mass coming from the Weyl anomaly so that,
\begin{equation}
\frac{2M_{a}}{g^{2}}\sim\frac{1}{2}F^{A}\partial_{A}f_{a}-\frac{c_{a}}{16\pi^{2}}F^{A}\hat{K}_{A}\label{eq:MWeyl}
\end{equation}
If we also perform a field redefinition to get canonical normalization
we have another term $\Delta f=\frac{1}{4\pi^{2}}\sum_{r}T(r)\tau_{Z}^{(r)}$
where the transformation parameter is fixed by $\tau_{Z}^{(r)}+\bar{\tau}_{Z}^{(r)}={\rm tr}\ln{\bf Z}^{(r)}|_{harm}.$
Here the sum $r$ is over the matter representations, and the matter
metric $Z{}^{(r)}$ is defined in \eqref{eq:Kexpn}. Thus modulo the
effect of redefining the gauge field kinetic term (see \citep{deAlwis:2008aq}
and Appendix) the expression for the gaugino mass becomes (for the
complete expression see \eqref{eq:mmu} of Appendix) 
\[
\frac{2M_{a}}{g_{{\rm phys}}^{2}}(\Phi,\bar{\Phi})=\frac{1}{2}F^{A}\partial_{A}f_{a}-\frac{c_{a}}{16\pi^{2}}F^{A}\hat{K}_{A}-\sum_{r}\frac{T_{a}(r)}{8\pi^{2}}F^{A}\partial_{A}{\rm tr}\ln{\bf Z}^{(r)}|_{0}.
\]
It is important to note that the coefficient of the second term is
the anomaly coefficient \eqref{eq:ca} rather than the beta function
coefficient \eqref{eq:beta}.

Note that this derivation (discussed in detail in Appendix) does not
make use of compensators at all. The alternative is to use a manifestly
Weyl invariant formalism - which is the form in which this contribution
was originally discussed in \citep{Kaplunovsky:1994fg}. In this case
instead of doing (an anomalous) field redefinition transformation,
one starts with the Weyl invariant formalism (reviewed below) and
adds a term which cancels the anomaly.

\subsection{The different compensators\label{sub:The-different-compensators}}

The most transparent way to identify the confusion in the AMSB literature
is to begin with the Weyl (chiral scalar) compensator formalism. 

The action given in the original paper \citep{Cremmer:1982en} (or
the superspace version of it given by Wess and Bagger \citep{Wess:1992cp})
does not reflect the (super) Weyl invariance of the torsion constraints
of supergravity. An action which is manifestly Weyl invariant (see
for example \citep{Kaplunovsky:1994fg} and references therein) is
the following (with $\kappa=M_{P}^{-1}=1,\, d^{8}z=d^{4}xd^{4}\theta,\, d^{6}z=d^{4}xd^{2}\theta$):
\begin{eqnarray}
S & = & -3\int d^{8}z{\bf E}C\bar{C}\exp[-\frac{1}{3}K(\Phi,\bar{\Phi};Q,\bar{Q}e^{2V})]+\nonumber \\
 &  & \left(\int d^{8}z{\cal E}[C^{3}W(\Phi,Q)+\frac{1}{4}f(\Phi,Q)){\cal W}{\cal W}]+h.c.\right).\label{eq:action-1}
\end{eqnarray}
In the above we've changed notation slightly so that the fields $\Phi$,
$Q$ are respectively a set of neutral (for example the moduli of
string theory) chiral superfields and ones charged under the gauge
groups. $V$ is the gauge prepotential and ${\bf {\cal W}}{}_{\alpha}=(-\frac{\bar{{\bf \nabla}}^{2}}{4}+2{\bf R})e^{-2V}\nabla_{\alpha}e^{2V}$
is the associated gauge field strength (for simplicity we have explicitly
displayed only one gauge group) and $\nabla_{\alpha}$is the covariant
super derivative. Also each term containing non-singlets is implicitly
taken to be an invariant. ${\bf R}$ is the chiral curvature superfield,
${\bf E}$ is the full superspace measure and ${\cal E}\equiv{\bf E}/2{\bf R}$
is the chiral superspace measure. The torsion constraints of SUGRA
are invariant under Weyl transformations (with a chiral superfield
transformation parameter $\tau$) which are given below.
\begin{eqnarray}
{\bf E}\rightarrow e^{2(\tau+\bar{\tau})}{\bf E}, &  & {\cal E}\,(\equiv\frac{{\bf E}}{2{\bf R}})\rightarrow e^{6\tau}({\cal E}+O(-\frac{{\bf \bar{\nabla}^{2}\bar{\tau}}}{4{\bf R}})),\label{eq:conf1}\\
\nabla_{\alpha}\rightarrow e^{(\tau-2\bar{\tau})}(\nabla_{\alpha}-2(\nabla^{\beta}\tau)M_{\beta\alpha}), &  & V\rightarrow V,\label{eq:conf2}\\
2{\bf R} & \rightarrow & e^{-4\tau}(-\frac{{\bf \bar{\nabla}^{2}}}{4}+2{\bf R)}e^{2\bar{\tau}}\label{eq:conf3}\\
\Phi\rightarrow\Phi,\, Q\rightarrow Q, &  & {\cal W}_{\alpha}\rightarrow e^{-3\tau}{\cal W}_{\alpha}.\label{eq:conf4}
\end{eqnarray}
Here $M_{\alpha\beta}$ is a Lorentz matrix. The chiral \textit{scalar}
Weyl compensator superfield $C$ with the transformation rule
\begin{equation}
C\rightarrow e^{-2\tau}C,\label{eq:Ctrans-1}
\end{equation}
is introduced in order to have a manifestly Weyl invariant action.
It can be hardly overemphasized that $C$ does not contain propagating
(i.e. physical) degrees of freedom. Clearly it can be gauged away
to unity in the classical action. Quantum mechanically these transformations
have an anomaly since the path integral measure is not invariant.
This anomaly was calculated in \citep{Kaplunovsky:1994fg} and gives
our equation \eqref{eq:ftransWeyl}. As we review below this is dealt
with by adding an appropriate correction to the gauge kinetic term
\citep{Kaplunovsky:1994fg}.

Note that ${\cal E}$ is a chiral measure appropriate for integrating
chiral scalars. In the chiral representation these integrals can be
reduced to chiral superspace integrals ($\int d^{6}z$). From the
superspace integration by parts rule and the torsion constraints \citep{Gates:1983nr,Buchbinder:1998qv}
the entire integral can be written effectively in the same form as
the last line of \eqref{eq:action-1}, since 
\[
\int d^{8}z{\bf E}{\bf L}=\int d^{8}z\frac{{\bf E}}{2{\bf R}}(-\frac{1}{4}\bar{{\bf {\bf \nabla}}}{}^{2}+2{\bf R){\bf L},}
\]
where ${\bf L}$ is an arbitrary unconstrained superfield. This relation
enables one to derive superspace equations of motion. Thus varying
the action w.r.t. $C$ gives 
\begin{equation}
-(-\frac{1}{4}\bar{{\bf {\bf \nabla}}}{}^{2}+2{\bf R)}(\bar{C}e^{-K/3})+C^{2}W=0\label{eq:Ceqn}
\end{equation}
In the $C=1$ gauge this equation becomes the trace equations that
are obtained by varying with respect to the conformal mode of the
super metric which is effectively $\delta_{\tau}S=0$. 

Taking the lowest component of \eqref{eq:Ceqn} gives (with $|_{0}$
an instruction to take the lowest component) 
\begin{equation}
\frac{\bar{F}^{\bar{C}}}{\bar{C}}+2{\bf R}|_{0}=\frac{1}{3}\bar{F}^{\bar{i}}K_{\bar{i}}|_{0}+\frac{C^{2}}{\bar{C}}e^{K/3}W|_{0}+{\rm fermion}\,{\rm terms}\label{eq:FC}
\end{equation}
In this compensator framework the Weyl anomaly needs to be cancelled
and KL do this by making the \emph{replacement} (compare with \eqref{eq:ftransWeyl}
which is the effect of a Weyl \emph{transformation} on the measure)
\begin{equation}
f_{a}(\Phi,Q)\rightarrow\tilde{f}_{a}(\Phi,Q)\equiv f_{a}(\Phi,Q)-\frac{3c_{a}}{8\pi^{2}}\ln C.\label{eq:f+lnC}
\end{equation}
 Next the field redefinition necessary to get to canonical normalization
for the matter terms needs to be done. We expand to lowest order in
the `MSSM' fields $Q$ and ignore higher than quadratic terms in these
fields since they are expected to get negligible vev's - see \eqref{eq:Kexpn}).
 This gives an additional term $-(T_{a}(r)\tau_{Z}/4\pi^{2}$, except
that now since we still have the compensator $C$ in the action, the
fixing of $\tau_{Z}$ is modified (compared to eqn \eqref{eq:konishifix}
of the Appendix) to%
\footnote{for simplicity we just consider one matter representation $r$.%
}
\begin{equation}
\tau_{Z}+\bar{\tau}_{Z}=\ln(C\bar{C}e^{-\hat{K}/3}\det Z_{r})|_{H}\label{eq:tauZfix}
\end{equation}
Here the instruction on the RHS requires one to keep just the harmonic
part of the expansion of $K$ (i.e. the sum of the chiral and anti-chiral
parts). Combined with \eqref{eq:f+lnC} this gives (apart from a term
coming from rescaling the gauge kinetic term - see Appendix) the quantum
gauge coupling function at the UV scale%
\footnote{All this is of course essentially contained in \citep{Kaplunovsky:1994fg}.
The explanation of how the anomaly coefficient $c_{a}$ gets replaced
by the beta function coefficient $b_{a}$ was given in \citep{deAlwis:2008aq}. %
},
\begin{eqnarray}
f_{a}^{quantum}+\bar{f}_{a}^{quantum} & \equiv\nonumber \\
f_{a}(\Phi,Q)+\bar{f}_{a}(\bar{\Phi},\bar{Q}) & - & \frac{3c_{a}}{8\pi^{2}}\ln(C\bar{C})-\frac{T_{a}(r)}{4\pi^{2}}\ln(C\bar{C}e^{-K_{m}/3}\det Z_{r})|_{H}\\
 & = & f_{a}(\Phi,Q)+\bar{f}_{a}(\bar{\Phi},\bar{Q})-\frac{b_{a}}{8\pi^{2}}\ln(C\bar{C})-\frac{T_{a}(r)}{4\pi^{2}}\ln(e^{-K_{m}/3}\det Z_{r})|_{H}\label{eq:fquantum}
\end{eqnarray}
This is our starting point for elucidating the confusion that exists
in the literature.

We note in passing that this scalar compensator formulation is separately
invariant under the Weyl transformations \eqref{eq:conf4} as well
as the Kähler transformation
\begin{eqnarray}
K(\Phi,\bar{\Phi};Q,\bar{Q}e^{2V}) & \rightarrow & K(\Phi,\bar{\Phi};Q,\bar{Q}e^{2V})+J(\Phi,Q)+\bar{J}(\bar{\Phi},\bar{Q}),\nonumber \\
W(\Phi,Q) & \rightarrow & e^{-J(\Phi,Q)}W(\Phi,Q)\nonumber \\
C & \rightarrow & e^{J(\Phi,Q)/3}C\nonumber \\
f_{a}(\Phi,Q) & \rightarrow & f_{a}(\Phi,Q)+\frac{c_{a}}{8\pi^{2}}J(\Phi,Q)\label{eq:Ktrans}
\end{eqnarray}

The correct gauge fixing of $C$ to get to the Einstein-Kähler frame
is \citep{Kaplunovsky:1994fg}
\begin{equation}
\ln C+\ln\bar{C}=\frac{K}{3}|_{H}.\label{eq:Cgaugefix}
\end{equation}
This amounts to going to the Wess-Zumino gauge for the real superfield
$K$. Needless to say this superfield gauge fixing is completely equivalent
to the set of transformations done in Wess and Bagger \citep{Wess:1992cp}
to get to the Einstein-Kähler frame. Thus for instance using this
gives from \eqref{eq:FC}
\begin{equation}
2{\bf R}|_{0}=e^{K/2}W|_{0}\equiv m_{3/2}.\label{eq:R0}
\end{equation}

\subsubsection{The correct calculation of the gaugino mass}

By using the gauge fixing of $C$ which is appropriate for getting
the Einstein-Kähler gauge (i.e. \eqref{eq:Cgaugefix}) we get for
the quantum gauge coupling function in this gauge the expression at
the UV scale%
\footnote{The full expression including the RG running term at some IR scale
is given in the Appendix.%
} 
\begin{equation}
\frac{1}{g_{a}^{2}(\Phi,\bar{\Phi})}=\Re f_{a}^{quantum}=\Re f_{a}(\Phi)-\frac{c_{a}}{16\pi^{2}}\ln K_{m}-\frac{T_{a}(r)}{8\pi^{2}}(\ln\det Z_{r})\label{eq:fquantumEgauge}
\end{equation}
The effective gauge coupling is then obtained by projecting the F-term
of this equation. This gives then the KL expression for the gaugino
mass 
\begin{equation}
\frac{2M_{a}}{g_{a}^{2}}(\Phi,\bar{\Phi};\mu)=\frac{1}{2}F^{A}\partial_{A}f_{a}-\frac{c_{a}}{16\pi^{2}}F^{A}\hat{K}_{A}-\sum_{r}\frac{T_{a}(r)}{8\pi^{2}}F^{A}\partial_{A}{\rm tr}\ln{\bf Z}^{(r)}(g^{2}),\label{eq:MquantumEgauge}
\end{equation}
This is of course in complete agreement with the alternative method
of calculation where instead of canceling the anomaly, one simply
picks up the extra terms from the anomaly upon doing the appropriate
transformations to get to the Einstein-Kähler gauge as discussed in
Appendix. Indeed the latter is how the Konishi terms are obtained
anyway in both methods!

\subsection{Misidentification of compensators and an incorrect calculation\label{sub:Misidentification-of-compensator}}

All calculations in the literature which claim to get an additional
term proportional to $m_{3/2}\equiv e^{K/2}W|_{0}$, start from a
formalism which is actually in the $C=1$ gauge, but then in effect
proceed to misidentify the density compensator (defined below) with
the scalar compensator $C$ . 

The density compensator arises from writing the supervielbein determinant
as \citep{Gates:1983nr}\citep{Buchbinder:1998qv} 
\begin{equation}
{\bf E}=\phi\bar{\phi}\hat{E}({\bf H})\label{eq:E}
\end{equation}
Here ${\bf H}$ is a (real) prepotential for the supergravity multiplet
and $\phi$($\bar{\phi}$) is the chiral density compensator ($\nabla_{\alpha}\bar{\phi}=0$).
Under Weyl transformations \eqref{eq:conf4}
\begin{equation}
{\bf H}\rightarrow{\bf H},\,\phi\rightarrow e^{2\tau}\phi,\,\bar{\phi}\rightarrow e^{2\bar{\tau}}\bar{\phi}.\label{eq:Hphi}
\end{equation}
However, even though $\phi$ is chiral it is not a scalar. Under super-diffeomorphisms
it transforms as a scalar density. In fact in the chiral representation
$\phi^{3}={\cal E}$, the measure on chiral superspace. Furthermore
\begin{equation}
\phi|_{0}=e^{1/3},\,-\frac{1}{4}\nabla^{2}\phi|_{0}=e^{1/3}2{\bf R}|_{0},\label{eq: phicomponents}
\end{equation}
with $e$ being the usual vierbein determinant. To explain the misidentification
it is easiest to start with the KL relation \eqref{eq:fquantum} which
is valid in a general gauge. Now perform a Weyl transformation \eqref{eq:conf4}
(i.e. $C\rightarrow e^{-2\tau}C,$ etc.) with the transformation parameter
(instead of $\ln C$ as in \eqref{eq:Cgaugefix}) fixed by%
\footnote{Note that in the region of field space that is of interest namely
where $Q\sim0$) the distinction between $K$ and $\hat{K}$ is irrelevant.%
} 
\begin{equation}
2\tau+2\bar{\tau}=-\frac{1}{3}\hat{K}\label{eq:tauKhat}
\end{equation}
 This then gives us from \eqref{eq:fquantum} the expression
\[
f_{a}^{quantum}=f_{a}(\Phi)-\frac{b_{a}}{8\pi^{2}}\ln C-\frac{c_{a}}{8\pi^{2}}\ln\hat{K}|_{chiral}-\frac{T_{a}(r)}{4\pi^{2}}(\ln\det Z_{r})|_{chiral}.
\]
.Note however that this expression is not in the Einstein-Kähler frame
unless one sets $C=1$, in which case of course we recover the KL
expression. Instead however what is done in the literature is to identify
$C$ with the (inverse of the) density compensator%
\footnote{Note that it is $\phi^{-1}$ which has the same Weyl transformation
as $C$ and therefore the corresponding term will cancel the Weyl
anomaly just as $C$ does.%
} $\phi^{-1}$ giving (purportedly in the Einstein frame) the expression
\begin{equation}
f_{a}^{quantum}=f_{a}(\Phi)+\frac{b_{a}}{8\pi^{2}}\ln\phi-\frac{c_{a}}{8\pi^{2}}\ln\hat{K}|_{chiral}-\frac{T_{a}(r)}{4\pi^{2}}(\ln\det Z_{r})|_{Harm}.\label{eq:fphi}
\end{equation}
Taking the F-term of this and using \eqref{eq: phicomponents} and
Einstein frame equation \eqref{eq:R0} we have the formula for the
gaugino mass \citep{Bagger:1999rd} that is often used in the AMSB
literature, 
\begin{equation}
\frac{2M_{a}}{g_{a}^{2}}(\Phi,\bar{\Phi};\mu)=\frac{1}{2}F^{A}\partial_{A}f_{a}+\frac{b_{a}}{16\pi^{2}}m_{3/2}+\frac{c_{a}}{16\pi^{2}}F^{A}\hat{K}_{A}-\sum_{r}\frac{T_{a}(r)}{8\pi^{2}}F^{A}\partial_{A}{\rm tr}\ln{\bf Z}^{(r)}(g^{2})|_{0}.\label{eq:BMP}
\end{equation}
The problem with this argument is that, as we've emphasized $\phi$
is not a scalar but a scalar density. Hence although the expression
takes care of Weyl anomalies it does so at the cost of introducing
a (super) diffeomorphism anomaly. In addition one sees that if one
takes the lowest component of \eqref{eq:fphi} there is a term in
the expression for $1/g^{2}$ that is proportional to $\ln e$. This
of course vanishes in flat space but it transforms as the log of a
density, and hence we see the lowest component of the super space
diffeomorphism anomaly i.e. a space-time diffeomorphism anomaly.

All versions of phenomenological AMSB expressions for the gaugino
mass are essentially variants on this incorrect calculation%
\footnote{The most recent example of this is \citep{D'Eramo:2012qd}. These
authors actually quote the KL formula for $f$, namely \eqref{eq:f+lnC},
but then proceed to confuse the scalar compensator ($C$ in our notation)
of KL with the density compensator $\phi$. These authors also give
arguments for both the origin of $b_{a}$ from $c_{a}$ (see\eqref{eq:fquantum}
above) as well as for the $\ln1/g^{2}$ term ( the last term of \eqref{eq:gboundary}
and the denominator of the RHS of \eqref{eq:mboundary}), These issues
had also been discussed in our earlier paper (eqns (16) through (18)
and (29) through (31) of \citep{deAlwis:2008aq} i.e. reference {[}12{]}
of \citep{D'Eramo:2012qd}. %
}. The literature on this is briefly surveyed in the following sub-subsections.

\subsubsection{The original AMSB calculations}

In these calculations \citep{Randall:1998uk}\citep{Giudice:1998xp}
the RG evolution contribution to the coupling constant is modified
by inserting a factor of $\phi\bar{\phi}$ inside the logarithm, i.e.
\begin{equation}
\Delta\frac{1}{g^{2}}=\frac{b}{16\pi^{2}}\ln\frac{\mu^{2}}{\Lambda^{2}}\rightarrow\frac{b}{16\pi^{2}}\ln\frac{\mu^{2}\phi\bar{\phi}}{\Lambda^{2}}\label{eq:RSbetasubstitution}
\end{equation}
As justification for the insertion of this spurion (see eqn. (27)
of \citep{Randall:1998uk} for example) the authors invoke precisely
the KL replacement for the gauge coupling function, i.e. the first
two terms on RHS of \eqref{eq:fquantum}. However they then misidentify
the scalar compensator $C$ of KL%
\footnote{It should be noted that in KL the scalar compensator (our $C$) is
called $\varphi$. The density compensator and the formula \eqref{eq:E}
are not used in KL.%
} with the (inverse of the) density compensator $\phi$. As explained
above, this would then of lead to the extra $m_{3/2}$ term at the
cost of introducing a diffeomorphism anomaly. Also in this work an
attempt is made to redefine the density compensator by writing ${\cal E}=e\Phi^{3}+\ldots$
where $e$ is the (bosonic) vielbein determinant. However this is
not a supersymmetric relation%
\footnote{The lowest component of $\Phi$ is 1. The supertransformation of 1
is zero - not a fermion! Essentially RS treats $\Phi$ as a spurion
in which case this causes an explicit breaking of SUSY. Alternatively
as advocated here the correct SUSY version is to use $\phi$ the density
compensator. This gives supersymmetric expressions but at the price
of violating general covariance.%
}. The correct relation is ${\cal E}=\phi^{3}+\ldots$ , with the lowest
and highest components given by eqn \eqref{eq: phicomponents}.

\subsubsection{The Bagger Poppitz Moroi calculation \citep{Bagger:1999rd}}

This calculation in effect takes the AMSB term (the second term on
the RHS of eqn \eqref{eq:BMP}) `derived' in \citep{Randall:1998uk}\citep{Giudice:1998xp}
and adds to it the KL expression (the sum of the third and fourth
terms). So the simplest way to see how this is obtained is to follow
the argument given in subsection \eqref{sub:Misidentification-of-compensator}
above. However these authors use instead (a modified version) of the
non-local anomaly action of \citep{LopesCardoso:1993sq}.
\begin{eqnarray}
\Delta L & = & -\frac{g^{2}}{(16\pi)^{2}}\int d^{2}\theta2{\cal E}{\cal W}{\cal W}\frac{4}{\square}(-\frac{\bar{\nabla}^{2}}{4}+2R)\nonumber \\
 &  & \{b_{a}4\bar{R}+\frac{T_{a}(r)}{3}\nabla^{2}K+T_{a}(r)\nabla^{2}\ln Z_{r}\}+h.c.\label{eq:BMPaction}
\end{eqnarray}
This is a non-local action. Furthermore since the inverse Laplacian
is the flat space one it is not clear how this can possibly be generally
covariant. So it appears that although it has global supersymmetry
it does not have local supersymmetry. So any conclusion derived from
this would necessarily suffer from the same problem as that using
the density compensator discussed above. In fact this action will
give a non-local contribution to the gauge coupling function which
violates general covariance in the same way that the local argument
(to the get the AMSB term) gave a general covariance violating gauge
coupling term (see discussion after \eqref{eq:BMP}).

In \citep{Bagger:1999rd} it is stated that the effective action must
reflect separately the anomalies under separate Weyl and Kähler transformations
(unlike the one in \citep{LopesCardoso:1993sq}) and indeed theirs
does that. However it is non-local and not generally covariant. The
KL construction on the other hand is both local and satisfies (super)
general covariance. It also accounts for both the Kähler anomaly and
the Weyl anomaly.

\subsubsection{1PI vs Wilsonian}

Now the question may arise as to whether the extra AMSB term in the
gaugino mass (the second term of \eqref{eq:BMP}), even though absent
in the Wilsonian action, is present in the 1PI action. However the
1PI action is ill-defined for gauge theories - especially for non-Abelian
confining gauge theories. In fact the notion of a gluino (or for that
matter gluon) mass makes sense only in the Wilsonian sense at some
scale above the confinement scale - since below that scale we have
a theory of mesons and nucleons. The MSSM and its symmetry breaking
soft terms make sense only in the Wilsonian sense. 

The object of a theory of supersymmetry breaking is the calculation
of these SUSY breaking soft terms at energy scales just above the
Weak scale or TeV scale. This is typically done by taking the input
from some GUT scale (or string scale) theory and running those parameters
down to the TeV scale using the standard MSSM RG equations. Thus once
the UV theory gives the value of $K_{m},$Z and $f_{a}$ at the UV
scale the formulae \eqref{eq:fquantumEgauge}\eqref{eq:MquantumEgauge}
(or more accurately the formulae \eqref{eq:gboundary}\eqref{eq:mboundary})
should be used as initial values for the RG evolution. This is a completely
Wilsonian procedure.

In fact the incorrect calculations in subsection \eqref{sub:Misidentification-of-compensator}
are also clearly Wilsonian. They are incorrect simply because of the
misidentification of the compensators discussed there. Thus the difference
between the two has nothing to do with Wilsonian vs. 1PI actions.

\subsubsection{Background independence and boundary terms }

In \citep{Gripaios:2008rg} an argument was given for an IR contribution
to the gaugino mass in SUSY AdS. This calculation is irrelevant if
one uses the correct formulae since the impetus for this was the fact
that the incorrect formula (i.e. \eqref{eq:BMP}) fails to vanish
in the supersymmetric limit where $F=0$. The paper then goes on to
argue that the existence of a infra red contribution (coming from
their claim that the boundary of AdS breaks SUSY) to the gaugino mass
implies a similar contribution from UV physics but of the opposite
sign, so that the cancellation between the two resolves the paradox
that the AMSB contribution (the second term on the RHS of \eqref{eq:BMP})
breaks supersymmetry explicitly.

There are several problems with this argument. Firstly since the UV
contribution should be described by the Wilsonian effective action,
the argument would imply that the Wilsonian formula for the gaugino
mass has a term which does not fit the framework of the SUSY effective
action as given for example in Appendix G of \citep{Wess:1992cp}.
This would mean that the Wilsonian effective action at the two derivative
level cannot be put in the standard SUGRA form. In other words there
is an explicit violation of local supersymmetry at the level of the
Wilsonian action. This is of course the main problem with the AMSB
claim, that we've identified in the introduction . This problem has
nothing to do with the background as such. It is a question of constructing
locally supersymmetric Wilsonian actions. Unless one is claiming that
there is an anomaly in either SUSY and or general covariance this
claim cannot be valid. 

Also these calculations are tied to a particular background. In this
case the argument is made in the AdS SUSY background. Presumably (although
the authors have not shown this) if the background breaks SUSY but
is still AdS (like in the well-known LVS constructions) the IR subtraction
that \citep{Gripaios:2008rg} advocate would (presumably) be proportional
to the (negative) CC. Thus they appear to be suggesting that the formula
\eqref{eq:BMP} should be modified such that the AMSB term is really
proportional not to $m_{3/2}$ %
\footnote{Recall that we defined the field dependent gravitino mass $m_{3/2}\equiv e^{K/2}W$. %
} but to the peculiar combination $(m_{3/2}(\Phi,\bar{\Phi})-\sqrt{|V_{0}}|/\sqrt{3}M_{P})$
so that it vanishes exactly at an AdS minimum ($V_{0}=-3m_{3/2}^{2}|_{0}M_{P}^{2}$)
and gives the AMSB expression in flat space. 

Such a formula is obviously not background independent%
\footnote{One might think of making this background independent by replacing
the term $\sqrt{|V_{0}|}$ in the above formula by $\sqrt{|V}|$.
However it hard to imagine how such a singular term can arise in the
effective action.%
}. It violates the expectation that a supersymmetric action should
have the form given in \citep{Wess:1992cp}. The purpose of an effective
action is to obtain after the inclusion of quantum corrections a background
independent low energy action that can then be used to find the background.
The action itself should not be background dependent. It certainly
violates the logical structure set out in the introduction on the
construction of locally supersymmetric effective actions. The formula
given by KL on the other hand satisfies these necessary requirements
and in particular the gaugino mass vanishes if supersymmetry is not
spontaneously broken.

\subsubsection{String theory calculations}

Although calculating the anomaly needs regularization the anomaly
itself is finite. Since all the arguments above (both the correct
and the incorrect ones) depend only on the anomaly itself (about which
there is no dispute), the relevance of string theory or some other
UV completion of the SUGRA is unclear to this author. The only point
at which the UV completion of SUGRA is necessary is in determining
the form of $K$, $W$, and $f$, at what ever scale the UV theory
is supposed to replace the low energy theory.

Nevertheless there have been two attempts at calculating the AMSB
term in the gaugino mass with contradictory results. The earlier work
\citep{Antoniadis:2005xa} found no such contribution so we will not
comment on it further. On the other hand the authors of \citep{Conlon:2010qy}
claimed to have found evidence of such a term. So let us examine the
basis for this claim.

The latter authors calculate the correlation function of two gaugino
vertex operators and a vertex operator for the NS NS flux and find
a non-vanishing value%
\footnote{Such a vertex operator (in contrast to that for the $B$ field) has
explicit factors of the ambient space coordinate function on the world
sheet. As such it is not in the class of well defined vertex operators
that one usually deals with in string theory.%
}. However even if this calculation is correct it is hard to see how
this would imply the authors claim that this is evidence for a coupling
of the gauge field kinetic terms to the flux superpotential. The latter
is determined by a set of flux integers i.e. integrals over three
cycles of NSNS and RR fluxes in the compact manifold. Such an object
is determined by topology and does not fluctuate. A non-zero correlation
function of the form found by these authors (if indeed it is non-zero)
can however come from other sources such as the expression for the
holomorphic Kähler modulus in terms of the geometric Kähler modulus
which involves a dependence on the flux \citep{Jockers:2004yj}.

Furthermore it is unclear whether this correlation function gives
a contribution to the effective action at a generic point in field
space, as opposed to the origin, which is where the string calculation
was done. The reason is that the integral that they compute is sensitive
to whether or not an infra red cutoff exists. This is similar to the
claimed violation of non-renormalization theorems in global supersymmetry
\citep{West:1990rm,Jack:1990pd}. In fact these violations would go
away too if we work at a generic point in field space - where the
field acts as an infra-red cutoff%
\footnote{We will discuss this in detail in a separate publication dealing with
the relation between the 1PI and the Wilsonian effective action in
supersymmetric theories.%
}.

\subsection{The DS effect}

The effect identified in \citep{Dine:2007me} (DS) is not equivalent
to what is normally called AMSB - which as discussed earlier is the
additional term proportional to $m_{3/2}$ that is supposed to exist
compared to the KL formula for the gauge coupling. The DS effect explicitly
needs a Higgs field in order to generate an AMSB like effect. In other
words instead of arbitrarily adding a spurion term these authors had
a physical chiral field which is charged under the gauge group and
acquires a vacuum expectation value. However the point is that the
AMSB contributions is supposed to exist even in the absence of a Higgs
field. For instance pure super Yang-Mills theory coupled to SUGRA
and neutral chiral scalar fields (that break supersymmetry) is expected
to exhibit an AMSB contribution. 

In \citep{deAlwis:2008aq} this effect was identified as a threshold
effect much like the effect in gauge mediation where the intermediate
scale of the messengers gives a contribution to the gaugino mass as
discussed in \citep{Giudice:1997ni}. However if one regards the Weyl
anomaly contribution as an effect given at some high scale then evolving
down to low scales after including all threshold contributions should
account for the correct gaugino mass at say the TeV scale. For further
discussion see \citep{deAlwis:2008aq}.

\section{Acknowledgments}

I wish to thank Felix Brummer, Michele Cicoli, Joe Conlon, Oliver
DeWolfe, Arthur Hebecker, Jan Louis, Marcus Luty, Anshuman Maharanna,
Fernando Quevedo, Michael Ratz and Alexander Westphal for discussions,
and F. D'Eramo J. Thaler and Z. Thomas, for correspondence. I also
wish to thank the Newton Institute, Cambridge and the organizers of
the BSM workshop for hospitality during the completion of this work.
The research of SdA is partially supported by the United States Department
of Energy under grant DE-FG02-91-ER-40672.

\section*{Appendix: Supergravity formalism and the soft SUSY breaking terms}

The most general manifestly supersymmetric action for chiral scalar
fields $\Phi$ coupled to supergravity and gauge fields, when restricted
to no more than two derivatives and is Weyl invariant, was written
down in equation \eqref{eq:action-1}. Now the superspace action given
in Wess and Bagger \citep{Wess:1992cp} and Gates et al \citep{Gates:1983nr}
is essentially the same \textit{except that} the Weyl compensator
field $C$ has been (effectively) set equal to unity - clearly this
can always be done as one can see from \eqref{eq:f+lnC}. Note that
as emphasized in a previous paper \citep{deAlwis:2008aq} the physics
of this theory cannot in anyway depend on the gauge fixing of this
Weyl invariance. Whether or not one uses the formalism with an explicit
compensator field $C$ is simply a matter of convenience. In this
appendix we will work in the gauge $C=1$ and discuss how field redefinition
anomalies (coming from the non-invariance of the measure in the path
integral) will yield the correct formula modifying the original form
of the gauge coupling function. 

The action as written in the $C=1$ gauge is 

\begin{eqnarray}
S & = & -3\int d^{8}z{\cal E}(-\frac{\bar{\nabla}^{2}}{4}+2R)\exp[-\frac{1}{3}K(\Phi,\bar{\Phi};Q,\bar{Q}e^{2V})]+\nonumber \\
 &  & \left(\int d^{8}z{\cal E}[W(\Phi,Q)+\frac{1}{4}f(\Phi){\cal W}^{a}{\cal W}^{a}]+h.c.\right).\label{eq:actionWB}
\end{eqnarray}
Our claim is that any physical effect should be obtainable from this
action (in other words nothing physical can depend on having the superfield
$C$ around). This means that once the functional form of $K,W$ and
$f$ are given one should be able to read off the physical masses
and couplings of the theory (at the scale at which we expect these
forms to be valid) from the expression in component form for the above
action that is given in (for instance) Appendix G of \citep{Wess:1992cp}.

\subsection*{General Expressions for soft terms and RG invariance}

What is of most interest for us in the context of (low energy) SUSY
breaking is the boundary values of the soft masses and couplings,
which in the context of the MSSM will become the parameters of phenomenological
interest. The theory above has a set of gauge neutral fields $\Phi=\{\Phi^{A}\}$,
which in a string theory context for instance would be identified
as the moduli determining the size and shape of the internal 6D manifold
as well as the string coupling. In general we need to find the point
at which these are stabilized in a SUSY breaking fashion and is such
that none of the charged fields $Q=\{Q^{a}\}$ get a vacuum value.
If one finds such a minimum then the soft masses are obtained in the
following manner \citep{Kaplunovsky:1993rd}.

We expand the superpotential and the Kähler potential in powers of
the charged fields, i.e. we write
\begin{eqnarray}
W & = & \hat{W}(\Phi)+\frac{1}{2}\tilde{\mu}_{ab}(\Phi)Q^{a}Q^{b}+\frac{1}{6}\tilde{Y}_{abc}(\Phi)Q^{a}Q^{b}Q^{c}+\ldots,\label{eq:Wexpn}\\
K & = & \hat{K}(\Phi,\bar{\Phi})+Z_{a\bar{b}}(\Phi,\bar{\Phi})Q^{a}Q^{\bar{b}}+[X_{ab}(\Phi,\bar{\Phi})Q^{a}Q^{b}+h.c.]+\ldots\label{eq:Kexpn}
\end{eqnarray}
Then one may easily compute the soft masses from the well known expression
for the scalar potential in supergravity and get \citep{Kaplunovsky:1993rd}\citep{Brignole:1997dp},
(ignoring D-term contributions)
\begin{eqnarray}
(m^{2})_{a}^{a'} & \equiv & Z^{a'\bar{b}}m_{a\bar{b}}^{2}=\frac{1}{3}(2V+F^{A}F^{\bar{B}}\hat{K}_{A\bar{B}})\delta_{a}^{a'}-F^{A}F^{\bar{B}}R_{A\bar{B}a\bar{b}}Z^{a'\bar{b}}\label{eq:softmass}\\
 & = & \frac{1}{3}(2V+F^{A}F^{\bar{B}}\hat{K}_{A\bar{B}})\delta_{a}^{a'}-F^{A}F^{\bar{B}}\partial_{A}(Z^{a'\bar{b}}\partial_{\bar{B}}Z_{a\bar{b}}).\label{eq:softmass1}
\end{eqnarray}
Note that these are background independent and RG invariant formulae.
This means that at any scale the soft mass is obtained by evaluating
the above at a minimum of the potential with the functions $K$ etc
being chosen at that scale. Note also that these are coefficients
of the Wilsonian action.

Now while \eqref{eq:actionWB} is manifestly (off-shell) supersymmetric,
the component form after eliminating auxiliary fields only has on-shell
supersymmetry. In fact in arriving at the latter a series of (super)
Weyl transformations and field redefinitions of chiral multiplets
has been performed. This is necessary in order to get to the Einstein
frame for (super) gravity and the the Kähler normalization for the
chiral fields (with for instance the scalar field kinetic term being
of the form $K_{a\bar{b}}\partial_{\mu}\phi^{b}\partial^{\mu}\bar{\phi}^{\bar{b}}$).
Now in the quantum theory these transformations do not leave the measure
invariant and there is an anomaly. However as usual this anomaly just
changes the gauge coupling function (at the two derivative level)
and has no effect on the Kähler metric. Hence the above formula \eqref{eq:softmass}
remains valid in the quantum theory - assuming of course that the
appropriate Kähler potential (metric) is used. For instance the dilaton
component of the Kähler potential of the heterotic string is a term
of the form $K\sim-\ln(S+\bar{S})$. Due to string loop effects this
term gets changed to $K\sim-\ln(S+\bar{S}-\Delta(M,\bar{M})/16\pi^{2})$
\citep{Derendinger:1991hq}. This will obviously change the curvature
term in \eqref{eq:softmass} - but this has nothing to do with an
anomaly. Similar considerations apply to the expressions for the $\mu$,
$B\mu$ and $A$ terms given in \citep{Kaplunovsky:1993rd}\citep{Brignole:1997dp}. 

The gauge coupling function on the other hand does experience an anomaly
since the above field redefinitions give contributions to the measure
which are of the form $\exp\{\#\int\tau{\cal W}{\cal W}+h.c.)$ where
$\tau$ is the (chiral) superfield Weyl transformation parameter.
Let us look at this in more detail. The relevant transformations consist
of two sets of superfield transformations \citep{Kaplunovsky:1994fg}:
First we have the Weyl transformations \eqref{eq:conf4}, with the
transformation parameter $\tau$ being determined by the condition
\begin{equation}
2\tau+2\bar{\tau}=-\frac{\hat{K}(\Phi,\bar{\Phi})}{3}|_{H},\label{eq:weylgaugefix}
\end{equation}
by the need to go to the Einstein-Kähler gauge. These transformations
essentially amount to fixing the real superfield $\hat{K}$ to the
Wess-Zumino gauge. There is an additional transformation that is needed
to get canonical normalization to the charged chiral fields:
\[
Q^{a}\rightarrow U_{\, b}^{a}Q^{b}
\]
where $U_{\,\, b}^{a}$ is a chiral matrix which is fixed by the condition
(for simplicity we are suppressing gauge and representation labels)
\citep{Kaplunovsky:1994fg,ArkaniHamed:1997mj} 
\begin{equation}
{\bf U}\bar{{\bf U}}^{T}=\exp\ln{\bf Z}|_{harm}.\label{eq:Ufix}
\end{equation}
 Again this amounts to going to a WZ gauge for the real field $\ln Z$.
In addition in order to get canonical normalization for the gauge
kinetic term an additional transformation needs to be made - this
is given by \citep{ArkaniHamed:1997mj} $V\rightarrow e^{(\tau_{V}+\bar{\tau}_{V})/2}V,$
(where $V$is the pre-potential). The parameter superfield $\tau_{V}$
is fixed by 
\begin{equation}
e^{-2\Re\tau_{V}}\equiv\frac{1}{g_{phys}^{2}(\Phi,\bar{\Phi})}\simeq\Re f(\Phi)\label{eq:tauVfix}
\end{equation}
 where $g_{phys}^{2}(\Phi,\bar{\Phi})$ is the final (physical) coupling
superfield and the last relation is valid to leading order. 
\[
\]
All three of these transformations however are anomalous and hence
the gauge coupling function is effectively replaced by (defining $\tau_{Z}^{(r)}\equiv{\rm tr}\ln U^{(r)}$)
\begin{equation}
f(\Phi)\rightarrow f(\Phi)+\frac{3c}{4\pi^{2}}\tau+\frac{1}{4\pi^{2}}\sum_{r}T(r)\tau_{Z}^{(r)}+\frac{1}{4\pi^{2}}T(G)\tau_{V}\label{eq:ftransform}
\end{equation}
where these chiral transformation parameters are fixed (up to a phase)
by\eqref{eq:weylgaugefix} \eqref{eq:tauVfix}and (from \eqref{eq:Ufix})
\begin{equation}
\tau_{Z}^{(r)}+\bar{\tau}_{Z}^{(r)}={\rm tr}\ln{\bf Z}^{(r)}|_{harm}.\label{eq:konishifix}
\end{equation}
Also $c=\sum_{r}T(r)-T(G)$ with $T(r)$ being the trace of a generator
in the representation $r$ and $r=G$ denotes the adjoint representation. 

It should be noted that these relations imply that the components
of these parameters are determined by 
\begin{eqnarray}
\Re\tau|_{o}=\frac{1}{12}\hat{K}|_{0} & \nabla_{\alpha}\tau|_{0}=\frac{1}{6}\psi_{\alpha}^{A}\hat{K}_{A}|_{0} & -\frac{1}{4}\nabla^{2}\tau|_{0}=\frac{1}{6}F^{A}\hat{K}_{A}|_{0}-\frac{1}{24}\psi_{\alpha}^{A}\psi^{B\alpha}\hat{K}_{AB}|_{0}\label{eq:taucomponents}\\
\Re\tau_{Z}^{(r)}|_{o}=\frac{1}{2}{\rm tr}\ln{\bf Z}|_{0} & \nabla_{\alpha}\tau_{Z}^{(r)}|_{0}=\psi_{\alpha}^{A}\partial_{A}{\rm tr}\ln{\bf Z}|_{0}^{(r)} & \tau_{ZF}^{(r)}\equiv-\frac{1}{4}\nabla^{2}\tau_{Z}^{(r)}|_{0}=\label{eq:tauZcomponents}\\
 &  & F^{A}\partial_{A}{\rm tr}\ln{\bf Z}|_{0}+\frac{1}{24}\psi_{\alpha}^{A}\psi^{B\alpha}\partial_{A}\partial_{B}{\rm tr}\ln{\bf Z}|_{0},\nonumber 
\end{eqnarray}
with similar relations for the components of $\tau_{V}$. Thus the
components of these transformation parameter superfields are expressed
in terms of the original chiral fields $\Phi$. These are precisely
the transformations that are performed on the original fields to get
the final form of the action in Einstein-Kähler frame in Appendix
G of \citep{Wess:1992cp}. The only modification due to the anomaly
is the replacement of the original $f$ according to the formula \eqref{eq:ftransform},
with the components of $\tau,\tau_{Z},\tau_{V}$, fixed by \eqref{eq:taucomponents}\eqref{eq:tauZcomponents}.
In particular this means that the gauge coupling function $g_{phys}(\Phi,\bar{\Phi})$and
the gaugino mass $M(\Phi,\bar{\Phi})$ are given by 
\begin{eqnarray}
\frac{1}{g_{phys}^{2}} & = & \Re f-\frac{c}{16\pi^{2}}\hat{K}|_{0}-\sum_{r}\frac{T(r)}{8\pi^{2}}\ln\det{\bf Z}^{(r)}|_{0}+\frac{T(G)}{8\pi^{2}}\ln\frac{1}{g_{phys}^{2}},\label{eq:gboundary}\\
\frac{2M}{g_{{\rm phys}}^{2}} & = & (\frac{1}{2}F^{A}\partial_{A}f-\frac{c}{16\pi^{2}}F^{A}\hat{K}_{A}-\sum_{r}\frac{T_{a}(r)}{8\pi^{2}}F^{A}\partial_{A}\ln\det{\bf Z}^{(r)}|_{0})\times(1-\frac{T(G_{a})}{8\pi^{2}}g_{{\rm {\rm phys}}}^{(a)2})^{-1}.\label{eq:mboundary}
\end{eqnarray}

It should be stressed that the three formulae \eqref{eq:softmass}\eqref{eq:gboundary}\eqref{eq:mboundary}
for the soft mass (and analogous formulae for the $\mu,B\mu$ and
$A$ terms), the gauge coupling and gaugino mass, are all expressions
valid at whatever scale the explicit expressions for the functions
$K$ etc. are valid. Thus if $K$is obtained from string theory then
(after incorporating $\alpha'$ and string loop corrections) one would
expect these expressions to be valid at some point close to the string
scale. These formulae are then to be used as the boundary conditions
for renormalization group (RG) evolution. To one loop order the RG
evolved value of the coupling function and the gaugino mass at some
scale $\mu$ would be given in terms of the value at the ($\Phi$
independent) boundary scale $\Lambda$, by making the replacements
\begin{equation}
f\rightarrow f-(b/16\pi^{2})\ln(\Lambda^{2}/\mu^{2}),\label{eq:fRG}
\end{equation}
 and $ $$g_{phys}^{2-1}\rightarrow f$ on the RHS of the first equation.
Note that to this order the second equation is unchanged and in fact
the factor in parenthesis in last term on the RHS can be replaced
by unity. However the above formulae can actually be interpreted as
being valid at some IR scale $\mu$, if in addition to the replacement
for $f$, one replaces $g_{phys}^{2}\rightarrow g_{phys}^{2}(\mu^{2})$
and 
\begin{equation}
{\bf Z}^{(r)}(\Phi,\bar{\Phi})\rightarrow{\bf Z}^{(r)}(\Phi,\bar{\Phi};g^{2}(\mu)).\label{eq:ZRG}
\end{equation}
Thus we have the following formulae for the parameters at the infra
red RG scale $\mu$, which are expected to be valid to all orders
in the loop expansion (in some renormalization scheme):

\begin{eqnarray}
\frac{1}{g_{phys}^{2}}(\Phi,\bar{\Phi};\mu) & = & \Re f-\frac{b}{16\pi^{2}}\ln(\frac{\Lambda^{2}}{\mu^{2}})+\frac{c}{16\pi^{2}}\hat{K}|_{0}-\sum_{r}\frac{T(r)}{8\pi^{2}}{\rm tr}\ln{\bf Z}^{(r)}(g^{2}(\mu))|_{0}\nonumber \\
 &  & +\frac{T(G)}{8\pi^{2}}\ln\frac{1}{g_{phys}^{2}(\mu)}\label{eq:gmu}\\
\frac{2M}{g_{{\rm phys}}^{2}}(\Phi,\bar{\Phi};\mu) & = & (\frac{1}{2}F^{A}\partial_{A}f-\frac{c}{16\pi^{2}}F^{A}\hat{K}_{A}-\sum_{r}\frac{T_{a}(r)}{8\pi^{2}}F^{A}\partial_{A}{\rm tr}\ln{\bf Z}^{(r)}(g^{2}(\mu))|_{0})\nonumber \\
 &  & \times(1-\frac{T(G_{a})}{8\pi^{2}}g_{{\rm {\rm phys}}}^{(a)2}(\mu))^{-1}\label{eq:mmu}
\end{eqnarray}
The first of these equations is the integrated form of the NSVZ beta
function with the boundary condition (at $\mu=\Lambda$) fixed by
the KL supergravity correction (the third term above). 

\[
\]
 \bibliographystyle{apsrev} \bibliographystyle{apsrev}
\bibliography{myrefs}

\begin{thebibliography}{23}
\expandafter\ifx\csname natexlab\endcsname\relax\def\natexlab#1{#1}\fi
\expandafter\ifx\csname bibnamefont\endcsname\relax
  \def\bibnamefont#1{#1}\fi
\expandafter\ifx\csname bibfnamefont\endcsname\relax
  \def\bibfnamefont#1{#1}\fi
\expandafter\ifx\csname citenamefont\endcsname\relax
  \def\citenamefont#1{#1}\fi
\expandafter\ifx\csname url\endcsname\relax
  \def\url#1{\texttt{#1}}\fi
\expandafter\ifx\csname urlprefix\endcsname\relax\def\urlprefix{URL }\fi
\providecommand{\bibinfo}[2]{#2}
\providecommand{\eprint}[2][]{\url{#2}}

\bibitem[{\citenamefont{Wess and Bagger}(1992)}]{Wess:1992cp}
\bibinfo{author}{\bibfnamefont{J.}~\bibnamefont{Wess}} \bibnamefont{and}
  \bibinfo{author}{\bibfnamefont{J.}~\bibnamefont{Bagger}},
  \bibinfo{journal}{Supersymmetry and supergravity}  (\bibinfo{year}{1992}),
  \bibinfo{note}{princeton, USA: Univ. Pr. 259 p}.

\bibitem[{\citenamefont{Kaplunovsky and Louis}(1993)}]{Kaplunovsky:1993rd}
\bibinfo{author}{\bibfnamefont{V.~S.} \bibnamefont{Kaplunovsky}}
  \bibnamefont{and} \bibinfo{author}{\bibfnamefont{J.}~\bibnamefont{Louis}},
  \bibinfo{journal}{Phys. Lett.} \textbf{\bibinfo{volume}{B306}},
  \bibinfo{pages}{269} (\bibinfo{year}{1993}), \eprint{hep-th/9303040}.

\bibitem[{\citenamefont{Kaplunovsky and Louis}(1994)}]{Kaplunovsky:1994fg}
\bibinfo{author}{\bibfnamefont{V.}~\bibnamefont{Kaplunovsky}} \bibnamefont{and}
  \bibinfo{author}{\bibfnamefont{J.}~\bibnamefont{Louis}},
  \bibinfo{journal}{Nucl. Phys.} \textbf{\bibinfo{volume}{B422}},
  \bibinfo{pages}{57} (\bibinfo{year}{1994}), \eprint{hep-th/9402005}.

\bibitem[{\citenamefont{Randall and Sundrum}(1999)}]{Randall:1998uk}
\bibinfo{author}{\bibfnamefont{L.}~\bibnamefont{Randall}} \bibnamefont{and}
  \bibinfo{author}{\bibfnamefont{R.}~\bibnamefont{Sundrum}},
  \bibinfo{journal}{Nucl. Phys.} \textbf{\bibinfo{volume}{B557}},
  \bibinfo{pages}{79} (\bibinfo{year}{1999}), \eprint{hep-th/9810155}.

\bibitem[{\citenamefont{Giudice et~al.}(1998)\citenamefont{Giudice, Luty,
  Murayama, and Rattazzi}}]{Giudice:1998xp}
\bibinfo{author}{\bibfnamefont{G.~F.} \bibnamefont{Giudice}},
  \bibinfo{author}{\bibfnamefont{M.~A.} \bibnamefont{Luty}},
  \bibinfo{author}{\bibfnamefont{H.}~\bibnamefont{Murayama}}, \bibnamefont{and}
  \bibinfo{author}{\bibfnamefont{R.}~\bibnamefont{Rattazzi}},
  \bibinfo{journal}{JHEP} \textbf{\bibinfo{volume}{12}}, \bibinfo{pages}{027}
  (\bibinfo{year}{1998}), \eprint{hep-ph/9810442}.

\bibitem[{\citenamefont{Bagger et~al.}(2000)\citenamefont{Bagger, Moroi, and
  Poppitz}}]{Bagger:1999rd}
\bibinfo{author}{\bibfnamefont{J.~A.} \bibnamefont{Bagger}},
  \bibinfo{author}{\bibfnamefont{T.}~\bibnamefont{Moroi}}, \bibnamefont{and}
  \bibinfo{author}{\bibfnamefont{E.}~\bibnamefont{Poppitz}},
  \bibinfo{journal}{JHEP} \textbf{\bibinfo{volume}{04}}, \bibinfo{pages}{009}
  (\bibinfo{year}{2000}), \eprint{hep-th/9911029}.

\bibitem[{\citenamefont{Gripaios et~al.}(2009)\citenamefont{Gripaios, Kim,
  Rattazzi, Redi, and Scrucca}}]{Gripaios:2008rg}
\bibinfo{author}{\bibfnamefont{B.}~\bibnamefont{Gripaios}},
  \bibinfo{author}{\bibfnamefont{H.~D.} \bibnamefont{Kim}},
  \bibinfo{author}{\bibfnamefont{R.}~\bibnamefont{Rattazzi}},
  \bibinfo{author}{\bibfnamefont{M.}~\bibnamefont{Redi}}, \bibnamefont{and}
  \bibinfo{author}{\bibfnamefont{C.}~\bibnamefont{Scrucca}},
  \bibinfo{journal}{JHEP} \textbf{\bibinfo{volume}{0902}}, \bibinfo{pages}{043}
  (\bibinfo{year}{2009}), \eprint{0811.4504}.

\bibitem[{\citenamefont{Conlon et~al.}(2011)\citenamefont{Conlon, Goodsell, and
  Palti}}]{Conlon:2010qy}
\bibinfo{author}{\bibfnamefont{J.~P.} \bibnamefont{Conlon}},
  \bibinfo{author}{\bibfnamefont{M.}~\bibnamefont{Goodsell}}, \bibnamefont{and}
  \bibinfo{author}{\bibfnamefont{E.}~\bibnamefont{Palti}},
  \bibinfo{journal}{Fortsch.Phys.} \textbf{\bibinfo{volume}{59}},
  \bibinfo{pages}{5} (\bibinfo{year}{2011}), \eprint{1008.4361}.

\bibitem[{\citenamefont{D'Eramo et~al.}(2012)\citenamefont{D'Eramo, Thaler, and
  Thomas}}]{D'Eramo:2012qd}
\bibinfo{author}{\bibfnamefont{F.}~\bibnamefont{D'Eramo}},
  \bibinfo{author}{\bibfnamefont{J.}~\bibnamefont{Thaler}}, \bibnamefont{and}
  \bibinfo{author}{\bibfnamefont{Z.}~\bibnamefont{Thomas}}
  (\bibinfo{year}{2012}), \eprint{1202.1280}.

\bibitem[{\citenamefont{Gates et~al.}(1983)\citenamefont{Gates, Grisaru, Rocek,
  and Siegel}}]{Gates:1983nr}
\bibinfo{author}{\bibfnamefont{S.~J.} \bibnamefont{Gates}},
  \bibinfo{author}{\bibfnamefont{M.~T.} \bibnamefont{Grisaru}},
  \bibinfo{author}{\bibfnamefont{M.}~\bibnamefont{Rocek}}, \bibnamefont{and}
  \bibinfo{author}{\bibfnamefont{W.}~\bibnamefont{Siegel}},
  \bibinfo{journal}{Superspace, or one thousand and one lessons in
  supersymmetry - Front. Phys.} \textbf{\bibinfo{volume}{58}},
  \bibinfo{pages}{1} (\bibinfo{year}{1983}), \eprint{hep-th/0108200}.

\bibitem[{\citenamefont{de~Alwis}(2008)}]{deAlwis:2008aq}
\bibinfo{author}{\bibfnamefont{S.~P.} \bibnamefont{de~Alwis}},
  \bibinfo{journal}{Phys. Rev.} \textbf{\bibinfo{volume}{D77}},
  \bibinfo{pages}{105020} (\bibinfo{year}{2008}), \eprint{0801.0578}.

\bibitem[{\citenamefont{Cremmer et~al.}(1983)\citenamefont{Cremmer, Ferrara,
  Girardello, and Van~Proeyen}}]{Cremmer:1982en}
\bibinfo{author}{\bibfnamefont{E.}~\bibnamefont{Cremmer}},
  \bibinfo{author}{\bibfnamefont{S.}~\bibnamefont{Ferrara}},
  \bibinfo{author}{\bibfnamefont{L.}~\bibnamefont{Girardello}},
  \bibnamefont{and}
  \bibinfo{author}{\bibfnamefont{A.}~\bibnamefont{Van~Proeyen}},
  \bibinfo{journal}{Nucl.Phys.} \textbf{\bibinfo{volume}{B212}},
  \bibinfo{pages}{413} (\bibinfo{year}{1983}).

\bibitem[{\citenamefont{Buchbinder and Kuzenko}(1998)}]{Buchbinder:1998qv}
\bibinfo{author}{\bibfnamefont{I.~L.} \bibnamefont{Buchbinder}}
  \bibnamefont{and} \bibinfo{author}{\bibfnamefont{S.~M.}
  \bibnamefont{Kuzenko}}, \bibinfo{journal}{Ideas and methods of supersymmetry
  and supergravity: Or a walk through superspace}  (\bibinfo{year}{1998}),
  \bibinfo{note}{Bristol, UK: IOP}.

\bibitem[{\citenamefont{Lopes~Cardoso and Ovrut}(1994)}]{LopesCardoso:1993sq}
\bibinfo{author}{\bibfnamefont{G.}~\bibnamefont{Lopes~Cardoso}}
  \bibnamefont{and} \bibinfo{author}{\bibfnamefont{B.~A.} \bibnamefont{Ovrut}},
  \bibinfo{journal}{Nucl. Phys.} \textbf{\bibinfo{volume}{B418}},
  \bibinfo{pages}{535} (\bibinfo{year}{1994}), \eprint{hep-th/9308066}.

\bibitem[{\citenamefont{Antoniadis and Taylor}(2005)}]{Antoniadis:2005xa}
\bibinfo{author}{\bibfnamefont{I.}~\bibnamefont{Antoniadis}} \bibnamefont{and}
  \bibinfo{author}{\bibfnamefont{T.~R.} \bibnamefont{Taylor}},
  \bibinfo{journal}{Nucl. Phys.} \textbf{\bibinfo{volume}{B731}},
  \bibinfo{pages}{164} (\bibinfo{year}{2005}), \eprint{hep-th/0509048}.

\bibitem[{\citenamefont{Jockers and Louis}(2005)}]{Jockers:2004yj}
\bibinfo{author}{\bibfnamefont{H.}~\bibnamefont{Jockers}} \bibnamefont{and}
  \bibinfo{author}{\bibfnamefont{J.}~\bibnamefont{Louis}},
  \bibinfo{journal}{Nucl.Phys.} \textbf{\bibinfo{volume}{B705}},
  \bibinfo{pages}{167} (\bibinfo{year}{2005}), \eprint{hep-th/0409098}.

\bibitem[{\citenamefont{West}(1991)}]{West:1990rm}
\bibinfo{author}{\bibfnamefont{P.~C.} \bibnamefont{West}},
  \bibinfo{journal}{Phys.Lett.} \textbf{\bibinfo{volume}{B258}},
  \bibinfo{pages}{375} (\bibinfo{year}{1991}).

\bibitem[{\citenamefont{Jack et~al.}(1991)\citenamefont{Jack, Jones, and
  West}}]{Jack:1990pd}
\bibinfo{author}{\bibfnamefont{I.}~\bibnamefont{Jack}},
  \bibinfo{author}{\bibfnamefont{D.}~\bibnamefont{Jones}}, \bibnamefont{and}
  \bibinfo{author}{\bibfnamefont{P.~C.} \bibnamefont{West}},
  \bibinfo{journal}{Phys.Lett.} \textbf{\bibinfo{volume}{B258}},
  \bibinfo{pages}{382} (\bibinfo{year}{1991}).

\bibitem[{\citenamefont{Dine and Seiberg}(2007)}]{Dine:2007me}
\bibinfo{author}{\bibfnamefont{M.}~\bibnamefont{Dine}} \bibnamefont{and}
  \bibinfo{author}{\bibfnamefont{N.}~\bibnamefont{Seiberg}},
  \bibinfo{journal}{JHEP} \textbf{\bibinfo{volume}{03}}, \bibinfo{pages}{040}
  (\bibinfo{year}{2007}), \eprint{hep-th/0701023}.

\bibitem[{\citenamefont{Giudice and Rattazzi}(1998)}]{Giudice:1997ni}
\bibinfo{author}{\bibfnamefont{G.~F.} \bibnamefont{Giudice}} \bibnamefont{and}
  \bibinfo{author}{\bibfnamefont{R.}~\bibnamefont{Rattazzi}},
  \bibinfo{journal}{Nucl. Phys.} \textbf{\bibinfo{volume}{B511}},
  \bibinfo{pages}{25} (\bibinfo{year}{1998}), \eprint{hep-ph/9706540}.

\bibitem[{\citenamefont{Brignole et~al.}(1997)\citenamefont{Brignole, Ibanez,
  and Munoz}}]{Brignole:1997dp}
\bibinfo{author}{\bibfnamefont{A.}~\bibnamefont{Brignole}},
  \bibinfo{author}{\bibfnamefont{L.~E.} \bibnamefont{Ibanez}},
  \bibnamefont{and} \bibinfo{author}{\bibfnamefont{C.}~\bibnamefont{Munoz}}
  (\bibinfo{year}{1997}), \eprint{hep-ph/9707209}.

\bibitem[{\citenamefont{Derendinger et~al.}(1992)\citenamefont{Derendinger,
  Ferrara, Kounnas, and Zwirner}}]{Derendinger:1991hq}
\bibinfo{author}{\bibfnamefont{J.}~\bibnamefont{Derendinger}},
  \bibinfo{author}{\bibfnamefont{S.}~\bibnamefont{Ferrara}},
  \bibinfo{author}{\bibfnamefont{C.}~\bibnamefont{Kounnas}}, \bibnamefont{and}
  \bibinfo{author}{\bibfnamefont{F.}~\bibnamefont{Zwirner}},
  \bibinfo{journal}{Nucl.Phys.} \textbf{\bibinfo{volume}{B372}},
  \bibinfo{pages}{145} (\bibinfo{year}{1992}), \bibinfo{note}{revised version}.

\bibitem[{\citenamefont{Arkani-Hamed and Murayama}(2000)}]{ArkaniHamed:1997mj}
\bibinfo{author}{\bibfnamefont{N.}~\bibnamefont{Arkani-Hamed}}
  \bibnamefont{and} \bibinfo{author}{\bibfnamefont{H.}~\bibnamefont{Murayama}},
  \bibinfo{journal}{JHEP} \textbf{\bibinfo{volume}{06}}, \bibinfo{pages}{030}
  (\bibinfo{year}{2000}), \eprint{hep-th/9707133}.

\end{thebibliography}

\end{document}